\documentclass[12pt,a4paper]{article}
\usepackage[latin1]{inputenc}
\usepackage{epsf}
\usepackage[body={6.5 in,9.0in},left=1.0 in,top=1.0 in]{geometry}
\usepackage{amsmath}
\usepackage{amsfonts}
\usepackage{hyperref}
\usepackage{amssymb}
\usepackage{graphicx}
\usepackage{makeidx}
\usepackage{cite}
\begin{document}

\author{Subhankar Roy\footnote{Email: meetsubhankar @ gmail.com} and N. Nimai Singh\footnote{E-mail: nimai03 @ yahoo.com}\\ \\
Department of Physics, Gauhati University, Guwahati-781014, India }
\title{\textbf{Leptonic mixing matrix in terms of Cabibbo angle}.}
\maketitle
\abstract
We phenomenologically  build a neutrino mass matrix which obeys the
$\mu-\tau$ symmetry with only  two free parameters, Cabibbo angle
($\lambda$)
 and a flavour twister parameter($\eta$). For vanishing $\eta$, the
 model assumes
 TBM mixing. When $\eta=\lambda$, the model generates a solar angle
 and solar mass squared difference that
 coincide with the experimental best-fits.
 It motivates us to propose a mixing matrix in the neutrino sector,
 with $\theta_{12}< \sin^{-1}(1/\sqrt{3})$, $\theta_{13}=0$ and
 $\theta_{23}=\pi/4$. 
The corrections in $\theta_{13}$ and $\theta_{23}$ are conducted
following the breaking of $\mu-\tau$ symmetry,
 by choosing a proper unitary diagonalizing charged lepton matrix, and
 this ensures the fact that $\theta_{23}$ lies within the first
 octant.
 The Cabibbo angle ($\lambda=\sin\theta_c$) plays the role of a
 guiding parameter
 in both neutrino as well as leptonic sector. The effect of the  Dirac
 CP violating phase
 $\delta_{cp}$ is also studied when it enters either through the
 charged leptonic mixing matrix or through neutrino mixing matrix.

\section{Introduction}
Tri-bimaximal mixing (TBM)\cite{pfh} is one of the most popular kinds
of mixing pattern. 
TBM is experienced in different flavour symmetry groups like $A_4$
\cite{em1,em2,em3,em4,em5,em6,a1,a2,az,ba},
 $S_4$\cite{csl1,wg,fbl1,fbl2} and $\Delta(54)$\cite{tk1,tk2,ht}.
 But recent experimental data questions the validity of TBM mixing as the first approximation. 

To understan the unification of quarks and leptons is one of biggest
goals in particle physics. 
In the ``bottom-up'' approach some phenomenological relations between
quark and lepton mixing parameters
 are sought out. Starting from a simple parametrization of neutrino
 mass matrix following $\mu-\tau$ symmetry, we derive
 a new mixing scheme in the neutrino sector as
\begin{eqnarray}
\label{unu}
U_{\nu}(\lambda)=\begin{pmatrix}
 \sqrt{\frac{4+\lambda}{6+\lambda}}& \sqrt{\frac{2}{6+\lambda}}  & 0 \\ 
-\frac{1}{\sqrt{6+\lambda}} &\sqrt{\frac{4+\lambda}{2(6+\lambda)}}  & \frac{1}{\sqrt{2}} \\ 
\frac{1}{\sqrt{6+\lambda}} & -\sqrt{\frac{4+\lambda}{2(6+\lambda)}}  & \frac{1}{\sqrt{2}}
\end{pmatrix}, 
\end{eqnarray}
where, $\lambda$ is the Wolfenstein parameter \cite{wk}; $\lambda \sim\,0.2253\,\pm 0.0007$\cite{nk}.
$U_{\nu}(\lambda)$ predicts the atmospheric mixing angle $\theta_{23}$
as maximal and reactor angle $(\theta_{13})$ 
as zero. But in contrast to TBM mixing, $U_{\nu}(\lambda)$ predicts
solar angle $(\theta_{12})$ rather 
with lesser value than $\sin^{-1}(1/\sqrt{3})$,
\begin{eqnarray}
\label{s13}
\theta_{12}=\sin^{-1}(\sqrt{\frac{2}{6+\lambda}}).
\end{eqnarray}
For vanishing $\lambda$, $U_{\nu}(\lambda)$ converges to general TBM mixing.

The present experimental data neither supports vanishing
$\theta_{13}$\cite{dc,db,rn,tk,pt} 
nor maximal $\theta_{23}$. The mixing angle, $\theta_{23}$ is expected
to 
lie within the first octant\cite{gf,mg}. $U_{\nu}$ is formulated in
the
 basis where charged lepton mass matrix is diagonal and the charged
 lepton mixing
 matrix $U_{eL}$ is an identity matrix. Under this assumption
 $U_{PMNS}$ \cite{bp} is equated to $U_{\nu}$. 
In order to evade the inadequacy appearing in atmospheric and reactor
angle, 
we try to construct a charged lepton diagonalizing matrix $U_{eL}$. We
have the standard relation,
$U_{PMNS}=U^{\dagger}_{eL}.U_{\nu}=U^{\dagger}_{eL}.U^{c\,\dagger}_{\nu}.U_{TB}$. 
The charged lepton mixing matrix $U_{eL}$ has to be a  unitary matrix
rather
 than an identity matrix. Following the work done by King\cite{sk},
 we try to formulate a diagonalizing charged lepton mixing matrix, $U_{eL}$.

In this approach the Cabibbo angle $(\lambda\sim\sin\theta_{c})$ seeds
the whole parametrization 
in neutrino as well as charged lepton sector. Finally we shall
construct the $PMNS$ matrix,
 $U_{PMNS}=U_{eL}^{\dagger}.U_{\nu}$ and look into the mixing angles. 
We shall also  try to investigate the same by introducing Dirac CP
phase, $\delta_{cp}$ in the PMNS matrix.
 The CP phase can be introduced either from charged lepton or neutrino sector.

In the process of parametrization of the $\mu-\tau$ symmetric mass
matrix, we start
 with certain ansatz that the mass squared differences for both solar
 as well as atmospheric,
 can be expressed in terms of the Wolfenstein parameter
 $\lambda$. This ansatz 
is phenomenological and is the outcome of the analysis of global
neutrino data\cite{vl}. Thus we have
\begin{eqnarray}
\label{delm}
\Delta\,m_{sol}^{2}&=& \lambda^{4} \,\,eV^2,\nonumber\\
\Delta\,m_{atm}^{2}&\sim & 9 \lambda^{8} \,\,eV^2 .
\end{eqnarray}

\label{pmmt}\section{Parametrization of $\mu-\tau$ symmetric neutrino mass matrix}
In ref.\cite{csl2} the author proposed an exact symmetry of the
neutrino mass matrix
 under the interchange of second and third generation of neutrinos.
 This symmetry is often called called $\mu-\tau$ or $2-3$ symmetry. 
Under this symmetry, matrix element of the neutrino mass matrix is
invariant 
under the interchange of flavor basis 
vectors,$| \,e\,\rangle \longleftrightarrow | \,e\,\rangle$ and
 $| \,\mu\,\rangle \longleftrightarrow -| \,\tau\,\rangle$. 
The appearence of minus sign ensures the positive mixing angles. With
this symmetry
 we get $m_{e\mu}=-m_{e\tau}$ and $m_{\mu\mu}=m_{\tau\tau}$.
 
This symmetry is characterised by some significant features like
predictions of
 maximal atmospheric, vanishing reactor and arbitrary solar mixing
 angles. 
The solar angle can be controlled with the proper choice of parameters present in the neutrino mass matrix, 
\begin{eqnarray}
\label{mutau}
M_{\mu\tau}&=&\begin{pmatrix}
a & b  & -b \\ 
b &c  & d  \\ 
-b & d & c
\end{pmatrix} m_0,\\
\tan 2\theta_{12}&=&\frac{2\sqrt{2} b}{a-c+d}.
\end{eqnarray}

If the neutrinos follow normal hierarchy (NH) mass pattern, then out of
the three absolute neutrino masses,
 one can be approximated to zero. We start with certain $\mu-\tau$
 symmetric mass matrix
 that can give one mass eigenvalues as zero. We choose,
\begin{eqnarray}
\label{m0}  
 M_{0}=\begin{pmatrix}
\xi & 0 & 0 \\ 
0 & \frac{1}{2} & \frac{1}{2} \\ 
0 & \frac{1}{2} & \frac{1}{2}
\end{pmatrix}m_{0}.
 \end{eqnarray}
The mass eigenvalues are $m_{\nu1}=0$, $m_{\nu2}=\xi\,m_{0}$ and
$m_{\nu3}=m_0$.
 But unfortunately $M_{0}$ predicts vanishing $\theta_{12}$. $M_{0}$ is mended to $M_{TB}$,
\begin{eqnarray}
M_{TB}=\begin{pmatrix}
\xi & \xi & -\xi \\ 
\xi &\frac{1}{2}+\xi  & \frac{1}{2}-\xi  \\ 
-\xi & \frac{1}{2}-\xi  & \frac{1}{2}+\xi 
\end{pmatrix}m_{0},
\end{eqnarray} 
with mass eigenvalues as $m_{\nu1}=0$, $m_{\nu2}=3\,\xi m_{0}$ and
$m_{\nu3}=m_0$.
 $M_{TB}$ predicts $\theta_{12}$ as $\sin^{-1}(1/\sqrt{3})$ which is
in 
accordance with TBM mixing. In order to account for the deviation from
TBM mixing, 
we introduce flavour twister $\eta$\cite{nh,nns1}. With $\eta$,
 we parametrize $M_{\mu\tau} (\xi,\eta,m_{0})$,
\begin{eqnarray}
\label{mmutau}
M_{\mu\tau}=\begin{pmatrix}
\xi & \xi(1+\frac{\eta}{4})^{1/2} & -\xi(1+\frac{\eta}{4})^{1/2} \\ 
 \xi(1+\frac{\eta}{4})^{1/2} &\frac{1}{2}(1+\frac{\xi\eta}{2})+\xi &\frac{1}{2}(1-\frac{\xi\eta}{2})-\xi   \\ 
- \xi(1+\frac{\eta}{4})^{1/2} &\frac{1}{2}(1-\frac{\xi\eta}{2})-\xi & \frac{1}{2}(1+\frac{\xi\eta}{2})+\xi
\end{pmatrix}m_{0}.
\end{eqnarray}
The above mass matrix predicts
\begin{eqnarray}
\label{tan}
\tan 2\theta _{12}=2\sqrt{2}\,\frac{(4+\eta)^{1/2}}{2+\eta}.
\end{eqnarray}
We obtain the mass eigenvalues ,
\begin{eqnarray}
\label{me}
m_{\nu 1}=0,\quad m_{\nu 2}=\frac{\xi}{2} (6+\eta)m_{0} \quad \text{and}\quad m_{\nu 3}=m_{0}.
\end{eqnarray} 
So far what we have done from eq.(\ref{m0}) up to eq.(\ref{tan}), is
simple algebraic rearrangement 
of the parameters. One significant outcome of this arrangement is that
$\tan 2\theta_{12}$ 
is freed  from other two parameters and is expressible in terms of
$\eta$ only. 
But this parametrization will turn unnatural if we assume the
parameters $\xi$, $\eta$ 
and $m_{0}$ as independent ones. In principle, nature will assign
certain numbers 
to the observational parameters. Hence such freedom is not
expected. We then
 try to incorporate certain constraints in order to reduce the number
 of free parameters 
in $M_{\mu\tau}$ and also seek certain dominant parameter present in the same.

\begin{figure}
\begin{center}
\includegraphics[scale=1]{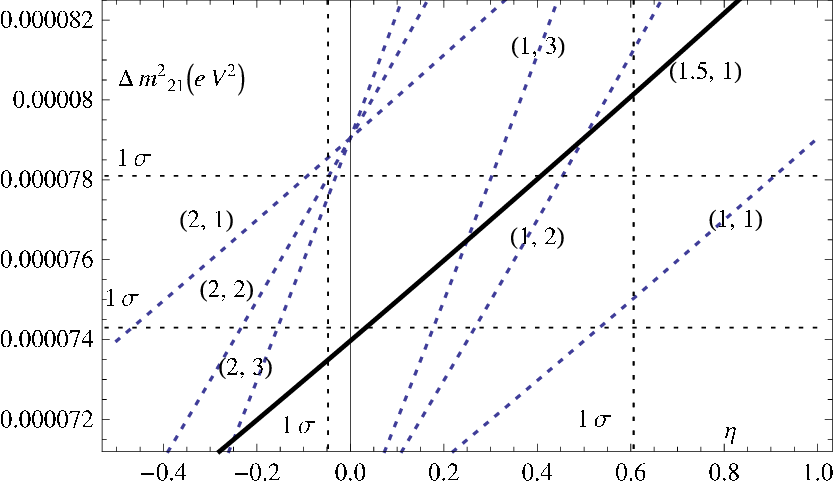} 
\caption{\footnotesize $m_{\nu2}$ is assumed as $m_{\mu3}(\lambda,\eta)$ with $3\lambda^{4}$ as the leading order. $\eta$ appears in the correction: $m_{\nu2}=3\lambda^{4}+(a_{0}+a_{1}\eta)\lambda^{5}$. The numerical values of $(a_{0},a_{1})$ is to be derived graphically. $\Delta m_{21}^{2}$ is plotted vs $\eta$, with different pairs of $(a_{0},a_{1})$. With respect to the $1\sigma$ range of $\eta$ derived with respect to $1\sigma$ range of solar angle data, this is preferable to choose the numerical value of $a_{1}$ as $1$ and $a_{0}$ in the range $1<a_{0}<2$.}
\end{center}
\end{figure}

The two ansatz we made in eq.(\ref{delm}), can be recast as
\begin{eqnarray}
\label{lam}
m_{\nu 3}&=& \lambda^2 \, eV,\nonumber \\
m_{\nu 2}& \sim & 3\lambda^4 \, eV=3\lambda^4 +\alpha \lambda^5\,eV,
\end{eqnarray}
where $\alpha$ is a correction parameter. With respect to $1\sigma$
range of $\Delta m_{21}^2$, 
we assign $\alpha$ a range $[1.533,1.908]$. From eq.(\ref{me}) and
eq.(\ref{lam}), 
we can directly infer : $m_{0}=\lambda^2$ and also the parameter $\xi$
depends on $\lambda$ and $\alpha$. 
We assume that $\alpha$ is not an independent parameter and it bears
some 
correlation with $\eta$. We shall check the relevance of this
assumption 
in the light of $1\sigma$ ranges of $\Delta m_{21}^2$ and $\eta$. 
$1\sigma$ range of solar angle confines $\eta$ within
$[-0.0476,0.606]$. 
We assume $\alpha$ to have a linear correlation with $\eta$; 
say $\alpha=\, a_{0}+a_{1}\eta $. With different pairs of $(a_0,a_1)$ 
we plot $\Delta m_{21}^2$ vs. $\eta$ in fig.1. In the perspective of
$1\sigma$ range, 
we find the suitable value of $a_{1}$ as $1$ and that for $a_{0}$,
$1<a_0<2$. 
We choose the average value of $a_{0}$ as $3/2$. Thus eq.(11) becomes

\begin{eqnarray}
\label{mn}
m_{\nu 2}=3\lambda^4 +(\frac{3}{2}+\eta) \lambda^5\,eV.
\end{eqnarray}  
Following eq.(\ref{me}), the above analysis leads to  
\begin{eqnarray}
\label{xi}
\xi=\, \xi (\lambda, \eta)=\frac{\lbrace 6+(\frac{3}{2}+\eta)\lambda\rbrace\lambda^{2}}{(6+\eta)}.
\end{eqnarray}
 Finally  $M_{\mu\tau}$ in eq.(\ref{mmutau}), is expressible with only
 two free parameters $\lambda$ and $\eta$ where 
  $\lambda=0.2253 \, \pm 0.0007$ and  $\eta$ is the flavour twister
 term which dominates in the expression of  $M_{\mu\tau}(\lambda,\eta)$.

\begin{figure}
\begin{center}
\includegraphics[scale=1]{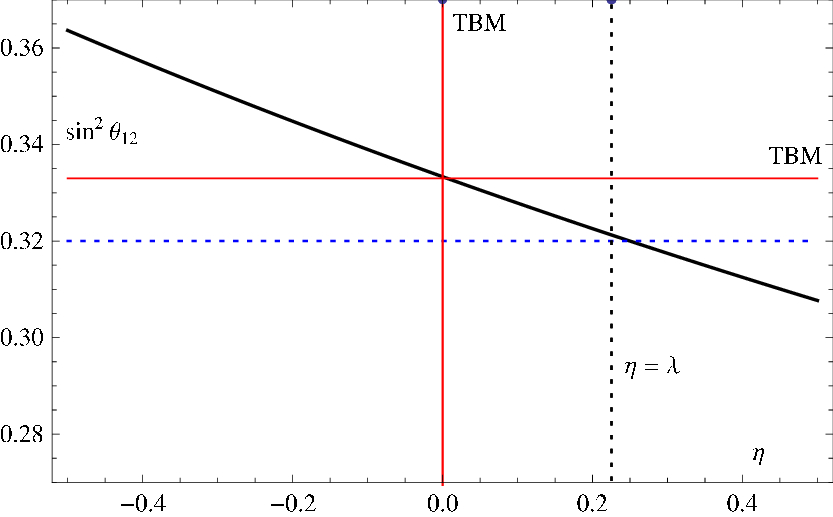} 
\caption{\footnotesize The solar angle is solely dictated by the flavour twister $\eta$. At $\eta=\lambda$, $\sin^{2}\theta_{12}$ is found to be $\sim 0.320$ (best-fit). $\sin^{2}\theta_{12}$ coincides with TBM prediction at $\eta=0$.}
\end{center}
\end{figure}

This is interesting to note that at $\eta=\lambda$, we obtain from eq.(\ref{tan}),
\begin{eqnarray}
\sin^{2}\theta_{12}=\frac{2}{6+\lambda}\approx \,0.321\; \text{(best-fit)}.
\end{eqnarray}
Again at  $\eta=\lambda$, we find
\begin{eqnarray}
\Delta m^2_{sol}=m_{\nu2}^2-m_{\nu1}^2 \approx \,7.6 \times10^{-5}\, eV^2\,\text{(best-fit)}.
\end{eqnarray}
TBM mixing is obtained at $\eta=0$, and the  corresponding mass
parameter,
 $\Delta m^2_{sol}=m_{\nu2}^2-m_{\nu1}^2 \approx \,7.4 \times10^{-5}\,
eV^2$ 
which is one of the boundaries of $1\sigma$ range.
The  simultaneous close agreement of the solar angle and solar mass
squared difference
 to the experimental best-fit at a single point thus hoists the
 preference for 
$\eta=\lambda$. $M_{\mu\tau}$ in eq.(\ref{mmutau}) with $\eta=\lambda$
 and $\xi=\xi(\lambda)$(eq.(\ref{xi})) generates the diagonalizing
 matrix $U_{\nu}(\lambda)$ in the exact form as shown in eq.(1),
\begin{eqnarray}
\label{unu}
U_{\nu}(\lambda)=\begin{pmatrix}
 \sqrt{\frac{4+\lambda}{6+\lambda}}& \sqrt{\frac{2}{6+\lambda}}  & 0 \\ 
-\frac{1}{\sqrt{6+\lambda}} &\sqrt{\frac{4+\lambda}{2(6+\lambda)}}  & \frac{1}{\sqrt{2}} \\ 
\frac{1}{\sqrt{6+\lambda}} & -\sqrt{\frac{4+\lambda}{2(6+\lambda)}}  & \frac{1}{\sqrt{2}}
\end{pmatrix}. 
\end{eqnarray}

\begin{figure}
\begin{center}
\includegraphics[scale=1]{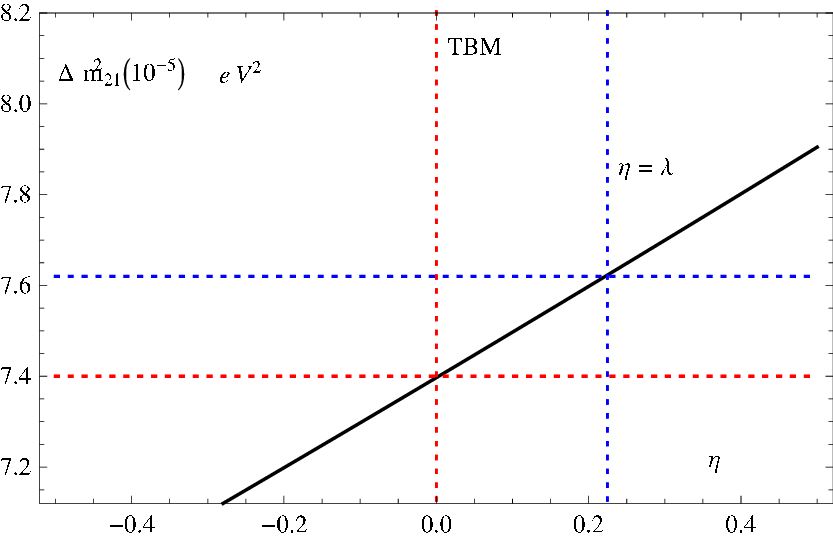} 
\caption{\footnotesize Choosing $\lambda=0.2253$, $\Delta m_{21}^{2}$ is plotted against $\eta$. It is  shown that at $\eta=\lambda$, $\Delta m_{21}^{2}\sim 7.62 \times 10^{-5}\, eV^{2}$ (best fit) is obtained.}
\end{center}
\end{figure}

We approximate $U_{\nu}(\lambda)$ as 
\begin{eqnarray}
U_{\nu}(\lambda)\approx\begin{pmatrix}
\sqrt{\frac{2}{3}}+\frac{\lambda}{12\sqrt{6}} & \frac{1}{\sqrt{3}}-\frac{\lambda}{12 \sqrt{3}} & 0 \\ 
-\frac{1}{\sqrt{6}}+\frac{\lambda}{12 \sqrt{6}} & \frac{1}{\sqrt{3}}+\frac{\lambda}{24 \sqrt{3}} & \frac{1}{\sqrt{2}} \\ 
 \frac{1}{\sqrt{6}}-\frac{\lambda}{12 \sqrt{6}}& -\frac{1}{\sqrt{3}}-\frac{\lambda}{24 \sqrt{3}} & \frac{1}{\sqrt{2}}
\end{pmatrix}. 
\end{eqnarray}
We see that for vanishing $\lambda$, $U_{\nu}(\lambda)$ converges to TBM mixing matrix, $U_{TB}$.
\begin{eqnarray}
U_{TB}=\begin{pmatrix}
\sqrt{\frac{2}{3}} & \frac{1}{\sqrt{3}} & 0 \\ 
-\frac{1}{\sqrt{6}} & \frac{1}{\sqrt{3}} & \frac{1}{\sqrt{2}} \\ 
 \frac{1}{\sqrt{6}}& -\frac{1}{\sqrt{3}} & \frac{1}{\sqrt{2}}
\end{pmatrix}.
\end{eqnarray}
This transition from $U_{TB}$ to $U_{\nu}(\lambda)$ can be interpreted in terms of an unitary matrix, $\tilde{U}=\,U_{TB}.U^{\dagger}_{\nu}$ and is displayed below,

\begin{eqnarray}
\label{uc}
\tilde{U}=\begin{pmatrix}
1 &\lambda/24 &-\lambda/24 \\ 
-\lambda/24 & 1 & 0\\ 
 \lambda/24 & 0  & 1 
\end{pmatrix}.
\end{eqnarray}
The unitary condition of $\tilde{U}$ is given as, 
$\tilde{U}^{\dagger}.\tilde{U}=\tilde{U}.\tilde{U}^{\dagger}=I+\mathcal{O}(10^{-5})$. 
Up-to this stage we are working on the the basis where charged lepton mass matrix is diagonal i.e., $U_{eL}=I$.

$M_{\mu\tau}(\lambda)$ which carries the information of absolute
neutrino masses and mixing angles can be recast as
\begin{eqnarray}
M_{\mu\tau}(\lambda)=\begin{pmatrix}
\lambda (1-\frac{\lambda}{6})& \lambda (1-\frac{\lambda}{24}) &-\lambda (1-\frac{\lambda}{24})\\ 
\lambda (1-\frac{\lambda}{24}) & \frac{1}{2}(1+\frac{5 \lambda^2}{36})+\lambda & \frac{1}{2}(1-\frac{7 \lambda^2}{36})-\lambda \\ 
-\lambda (1-\frac{\lambda}{24})& \frac{1}{2}(1-\frac{7 \lambda^2}{36})-\lambda &\frac{1}{2}(1+\frac{5 \lambda^2}{36})+\lambda
\end{pmatrix}\lambda^2 + \mathcal{O}(\lambda^5).
\end{eqnarray} 
In short, $M_{\mu\tau}$ predicts the solar angle, 
$\theta_{12}=34.51^{0}$ (deviated from TBM prediction) atmospheric
angle,
 $\theta_{23}=45^0$ and reactor angle, $\theta_{13}=0^0$. 
In order to account for the required deviations in these 
two mixing angles without disturbing the first, 
we  try to correct $M_{\mu\tau}(\lambda)$ with a proper 
choice of a unitary charged lepton diagonalizing matrix, $U_{eL}$. 
The corrected neutrino mass matrix is
$M_{\nu}=U_{eL}^{\dagger}.M_{\mu\tau}.
 U_{eL}$ is expected to give the complete picture of the mixing angles.

\label{clmm}\section{Charged lepton mixing matrix}
  In ref\cite{sk}, a new idea of mixing was proposed by King. This was
  based
 on the relationship\cite{cg,hm,ad,ll} consistent with the recent data. 
  \begin{eqnarray}
\sin\theta_{13}=\frac{\sin\theta_{c}}{\sqrt{2}}=\frac{\lambda}{\sqrt{2}}.
\end{eqnarray} 
The above ansatz implies $\theta_{13}\approx \,9.2^0$. The author
combined the above
 relation with TBM mixing leading to the proposal of the  Tri-bimaximal-Cabibbo mixing;
\begin{eqnarray}
\theta_{13}=\, \sin^{-1}(\lambda/\sqrt{2}), \quad \theta_{12}=\, \sin^{-1}(1/\sqrt{3}),\quad \theta_{23}=\pi/4.
\end{eqnarray}
With $\delta_{cp}=0$, the mixing matrix assumes the following texture.
\begin{equation}
\label{utbc}
U_{TBC}=
\begin{pmatrix}
\sqrt{\frac{2}{3}}(1-\frac{1}{4}\lambda^{2}) & \frac{1}{\sqrt{3}}(1-\frac{1}{4}\lambda^{2}) & \frac{1}{\sqrt{2}}\lambda \\ 
-\frac{1}{\sqrt{6}}(1+\lambda) &\frac{1}{\sqrt{3}}(1-\frac{1}{2}\lambda)  & \frac{1}{\sqrt{2}}(1-\frac{1}{4}\lambda^{2}) \\ 
\frac{1}{\sqrt{6}}(1-\lambda) &-\frac{1}{\sqrt{3}}(1+\frac{1}{2}\lambda)  & \frac{1}{\sqrt{2}}(1-\frac{1}{4}\lambda^{2}).
\end{pmatrix} +\mathcal{O}(\lambda^3).
\end{equation}
The above mixing scheme has its own limitations as it fails to lower the solar and atmospheric
mixing angles from TBM predictions. 
Also it ignores the preference for $\theta_{23}$ to lie within the
first octant. 
The significant feature of this mixing scheme lies in the prediction of a non-zero reactor angle. 

We can think $U_{TBC}$ as one of the corrected version of $U_{TB}$. 
We hope that certain choice of the charged lepton mixing matrix, say
$U_{eL}^{k}$ 
is responsible for the above
texture. $U_{eL}^{k}=U_{TB}U_{TBC}^{\dagger}$ where,

\begin{equation}
U_{eL}^{k}=\begin{pmatrix}
1-\frac{1}{4}\lambda^{2} &-\frac{1}{2}\lambda  &-\frac{1}{2}\lambda  \\ 
\frac{1}{2}\lambda & 1-\frac{1}{8}\lambda^{2} &-\frac{1}{8}\lambda^{2}  \\ 
 \frac{1}{2}\lambda &-\frac{1}{8}\lambda^{2}  & 1-\frac{1}{8}\lambda^{2}
\end{pmatrix}.
\end{equation}
The charged lepton mixing matrix satisfies the unitary condition,
 $U^{k\dagger}_{eL}U_{eL}^{k}=U_{eL}^{k}U^{k\dagger}_{eL}=I+\mathcal(O)(\lambda^4)$.
We mend $U_{eL}^k$ on associating it with $R_{23}(\theta)$\cite{nns2} and construct the new charged lepton mixing matrix,$U_{eL}$. Where, $U_{eL}=U_{eL}^k.R_{23}(\theta)$ and $R_{23}(\theta)$ is the rotational matrix in $2-3$ plane. 
\begin{eqnarray}
R_{23}=\begin{pmatrix}
1 & 0 & 0 \\ 
0 & 1-\frac{1}{2}\theta^{2}& \theta \\ 
0 & -\theta & 1-\frac{1}{2}\theta^{2}
\end{pmatrix}
\end{eqnarray}\label{f1}
With $\theta=\lambda/3$, we obtain, the required diagonalizing charged lepton mixing matrix 
\begin{eqnarray}
\label{uel}
U_{eL}=\begin{pmatrix}
1-\frac{\lambda^{2}}{4} &-\frac{1}{2}\lambda\lbrace 1-\frac{\lambda}{3}(1+\frac{\lambda}{6})\rbrace & -\frac{1}{2}\lambda\lbrace 1+\frac{\lambda}{3}(1-\frac{\lambda}{6})\rbrace \\ 
\frac{1}{2}\lambda & 1-\frac{\lambda^{2}}{24}(\frac{13}{3}-\lambda) &\frac{1}{3}\lambda-\frac{\lambda^{2}}{8}(1+\frac{\lambda}{3})\\ 
\frac{1}{2}\lambda &-\frac{1}{3}\lambda-\frac{\lambda^{2}}{8}(1-\frac{\lambda}{3}) & 1-\frac{\lambda^{2}}{24}(\frac{13}{3}+\lambda)
\end{pmatrix}+\mathcal{O}(\lambda^{4})
\end{eqnarray}

Upon incorporating $U_{eL}$ with the TBM mixing matrix, $U_{TB}$, $U'=U_{eL}^{\dagger}.U_{TB}$,
\begin{equation}
\label{un}
U'=\begin{pmatrix}
\sqrt{\frac{2}{3}}(1-\frac{\lambda^{2}}{4}) & \frac{1}{\sqrt{3}}(1-\frac{\lambda^{2}}{4}) & \frac{\lambda}{\sqrt{2}} \\ 
-\frac{1}{\sqrt{6}}\lbrace(1+\frac{4}{3}\lambda)-\frac{7\lambda^{2}}{18}(1+\frac{\lambda}{7})\rbrace &\frac{1}{\sqrt{3}}\lbrace(1-\frac{\lambda}{2})+\frac{\lambda^{2}}{9}(1+\frac{\lambda}{4}) \rbrace & \frac{1}{\sqrt{2}}(1-\frac{\lambda}{3})-\Lambda_{-} \\ 
\frac{1}{\sqrt{6}}\lbrace(1-\frac{4}{3}\lambda)-\frac{7\lambda^{2}}{18}(1-\frac{\lambda}{7})\rbrace & -\frac{1}{\sqrt{3}}\lbrace(1+\frac{\lambda}{2})+\frac{\lambda^{2}}{9}(1-\frac{\lambda}{4}) \rbrace & \frac{1}{\sqrt{2}}(1+\frac{\lambda}{3})-\Lambda_{+} .
\end{pmatrix},
\end{equation}
where, $
\Lambda{\pm}=\frac{11}{36\sqrt{2}}\lambda^{2}(1\pm \frac{3}{11}\lambda).
$

Similar to $U_{TBC}$ in eq.(\ref{utbc}), $U'$ predicts same
$\theta_{13}$ and $\theta_{12}$.
 But $U'$  generates a $\theta_{23}$ with lesser value than the maximal and of course $\theta_{23}$ lies within the first octant. 
\begin{eqnarray}
\sin^2\theta_{23}&\approx&\frac{1}{2}-\frac{\lambda}{3}(1-\frac{\lambda^2}{18})\approx 0.425.
\end{eqnarray}
This result is very close to $\sin^2\theta_{23}=0.427$ (best-fit).

\section{Perturbing the $\mu-\tau$ symmetry of neutrino mass matrix}
This is clear from the discussions in section (\ref{pmmt}) and section
(\ref{clmm}) that the solar angle can be suppressed within
 the neutrino sector whereas the lepton sector can control the
 atmospheric as well as
 the reactor angles. The corrected neutrino mass matrix with broken $\mu-\tau$ symmetry is,
\begin{align}
\label{mn}
M_{\nu}&=U_{eL}^{\dagger}.M_{\mu\tau}.U_{eL}\nonumber\\
 &= M^{\nu}_{\mu\tau}+\Delta M_{\nu},
\end{align}
where $\Delta M_{\nu}$ estimates the amount of perturbation to break
the $\mu-\tau$ symmetry for producing the necessary deviations in the
mixing angles. The diagonalizing matrix of $M_{\nu}(\lambda)$ has the
following texture,  $U_{PMNS}=U_{eL}^{\dagger}.U_{\nu}(\lambda)$,
\begin{align}
&U_{PMNS}=\nonumber\\
&\begin{pmatrix}
\sqrt{\frac{2}{3}}(1-\frac{\lambda^2}{4})(1+\frac{\lambda}{24})& \frac{1}{\sqrt{3}}(1-\frac{\lambda^2}{4})(1-\frac{\lambda}{12}) & \frac{\lambda}{\sqrt{2}} \\ 
-\frac{1}{\sqrt{6}}\lbrace 1+\frac{5}{4}\lambda (1-\frac{3}{10}\lambda)\rbrace +\frac{7}{\sqrt{2}}\Lambda_{0} & \frac{1}{\sqrt{3}}\lbrace 1-\frac{\lambda}{8}(1-\frac{4}{3}\lambda)\rbrace +\frac{5}{4}\Lambda_{0} & \frac{1}{\sqrt{2}}(1-\frac{\lambda}{3})-\Lambda_{-} \\ 
\frac{1}{\sqrt{6}}\lbrace 1-\frac{\lambda}{12}(17-\frac{29}{6}\lambda)\rbrace +\frac{5}{\sqrt{2}}\Lambda_{0} &-\frac{1}{\sqrt{3}}\lbrace 1+\frac{5}{24}\lambda (1+\frac{4}{15}\lambda)\rbrace +\frac{19}{4}\Lambda_{0} &\frac{1}{\sqrt{2}}(1+\frac{\lambda}{3})-\Lambda_{+} 
\end{pmatrix}, 
\end{align}
where, $\Lambda_{0}=\lambda^3/(108 \sqrt{3})$.

The mixing angles are given by the three relations,
\begin{eqnarray}
\label{eq1}
\sin^2\theta_{13}=|U_{e3}|^2,\quad \sin^2\theta_{12}=\frac{|U_{e2}|^2}{1-|U_{e3}|^2}, \quad \sin^2\theta_{23}=\frac{|U_{\mu3}|^2}{1-|U_{e3}|^2}.
\end{eqnarray}
We obtain,
\begin{eqnarray}
\label{s1}
\sin^2\theta_{13}&=&\lambda^2/2 \\
\label{s2} 
\sin^2\theta_{12}&\approx &\frac{1}{3}-\frac{\lambda}{18}(1-\frac{\lambda}{24}),\\ \label{s3}
\sin^2\theta_{23}&\approx&\frac{1}{2}-\frac{\lambda}{3}(1-\frac{\lambda^2}{18}).
\end{eqnarray}
The corresponding numerical values (with $\lambda=0.2253$) of the mixing angle parameters are $0.0253\,(\sin^2\theta_{13})$, $0.321\,(\sin^2\theta_{12})$ and $0.425\,(\sin^2\theta_{23})$.

\section{The effect of Dirac CP phase,$\delta_{cp}$}

The Dirac CP phase can enter $U_{PMNS}$, either through the charged
 lepton mixing matrix, $U_{eL}$ (lepton sector) or through $U_{\nu}$
 (neutrino sector).
 $U_{\nu}$ comprises $\tilde{U}$ and $U_{TB}$. Hence there are two
 possibilities by
 which $\delta_{cp}$ can enter the  neutrino sector i.e.,
 either through any of the two matrices.
\subsection{$\delta_{cp}$ from $U_{eL}$ (leptonic sector)}
We choose the texture of $U_{eL}$ with $\delta_{cp}$ in such a way that unitary condition is not violated,

\begin{equation}
U_{eL}(\lambda,\delta_{cp})=\begin{pmatrix}
1-\frac{\lambda^{2}}{4} &-\frac{1}{2}\lambda e^{-i \delta_{cp}}\lbrace 1-\frac{\lambda}{3}(1+\frac{\lambda}{6})\rbrace & -\frac{1}{2}\lambda e^{-i \delta_{cp}} \lbrace 1+\frac{\lambda}{3}(1-\frac{\lambda}{6})\rbrace \\ 
\frac{1}{2}\lambda e^{i\delta_{cp}} & 1-\frac{\lambda^{2}}{24}(\frac{13}{3}-\lambda) &\frac{1}{3}\lambda-\frac{\lambda^{2}}{8}(1+\frac{\lambda}{3})\\ 
\frac{1}{2}\lambda e^{i \delta_{cp}} &-\frac{1}{3}\lambda-\frac{\lambda^{2}}{8}(1-\frac{\lambda}{3}) & 1-\frac{\lambda^{2}}{24}(\frac{13}{3}+\lambda)
\end{pmatrix}+\mathcal{O}(\lambda^{4}).
\end{equation}

The elements present in the $U_{PMNS}=U_{eL}^{\dagger}.U_{\nu}$,
except $U_{e1}$, $U_{e2}$, $U_{\mu 2}$ and $U_{\tau3}$ 
 are affected by $\delta_{cp}$. But the predictions of the mixing
 angles are not affected by the presence 
of CP phase and these coincide with the predictions in eqs.(\ref{s1},\ref{s2},\ref{s3}).

\subsection{$\delta_{cp}$ from $U_{\nu}$ through $U_{TB}$ (neutrino sector)}

We choose  the TBM mixing matrix, $U_{TB}$ ,
\begin{align}
U_{TB}(\lambda,\delta_{cp})&=\begin{pmatrix}
\sqrt{\frac{2}{3}} & \frac{1}{\sqrt{3}} e^{-i \delta_{cp}} & 0 \\ 
-\frac{1}{\sqrt{3}}e^{i \delta_{cp}} & \sqrt{\frac{2}{3}} & 0  \\ 
0 & 0 & 1
\end{pmatrix}. R_{23}(\pi/4) \nonumber \\
&=\begin{pmatrix}
\sqrt{\frac{2}{3}} &\frac{1}{\sqrt{3}}e^{-i \delta_{cp}}  & 0 \\ 
-\frac{1}{\sqrt{6}}e^{i \delta_{cp}} & \frac{1}{\sqrt{3}} & \frac{1}{\sqrt{2}} \\ 
\frac{1}{\sqrt{6}}e^{i \delta_{cp}} & -\frac{1}{\sqrt{3}} & \frac{1}{\sqrt{2}}
\end{pmatrix},
\end{align}
and obtain $U_{\nu}(\lambda,\delta_{cp})$,
\begin{eqnarray}
U_{\nu}(\lambda,\delta_{cp})=\begin{pmatrix}
\sqrt{\frac{2}{3}}+\frac{\lambda e^{i \delta_{cp}}}{12 \sqrt{6}} & -\frac{\lambda}{12 \sqrt{3}}+\frac{e^{-i \delta_{cp}}}{\sqrt{3}} & 0 \\ 
\frac{\lambda}{12 \sqrt{6}}-\frac{e^{i \delta_{cp}}}{\sqrt{6}} &\frac{1}{\sqrt{3}}+\frac{\lambda e^{-i \delta_{cp}}}{24 \sqrt{3}}  & \frac{1}{\sqrt{2}} \\ 
-\frac{\lambda}{12 \sqrt{6}}+\frac{e^{i \delta_{cp}}}{\sqrt{6}} & -\frac{1}{\sqrt{3}}-\frac{\lambda e^{-i \delta_{cp}}}{24 \sqrt{3}} & \frac{1}{\sqrt{2}}
\end{pmatrix}.
\end{eqnarray}
Upon investigating the elements in PMNS matrix,
$U_{PMNS}=U_{eL}^{\dagger}(\lambda).U_{\nu}$, we find only $|U_{e2}|$
to be affected by $\delta_{CP}$. The prediction of $\sin^2\theta_{12}$
is affected while the rest preserves
 the same expression given in eqs.(\ref{s1},\ref{s3}),
\begin{eqnarray}
\label{s11}
\sin^{2}\theta_{12}=\frac{1}{3}-\frac{\lambda}{18}(\cos\delta_{cp}-\frac{\lambda}{24}).
\end{eqnarray}

\subsection{$\delta_{cp}$ from $U_{\nu}$ through $\tilde{U}$}
\begin{figure}
\begin{center}
\label{dd}
\includegraphics[scale=1]{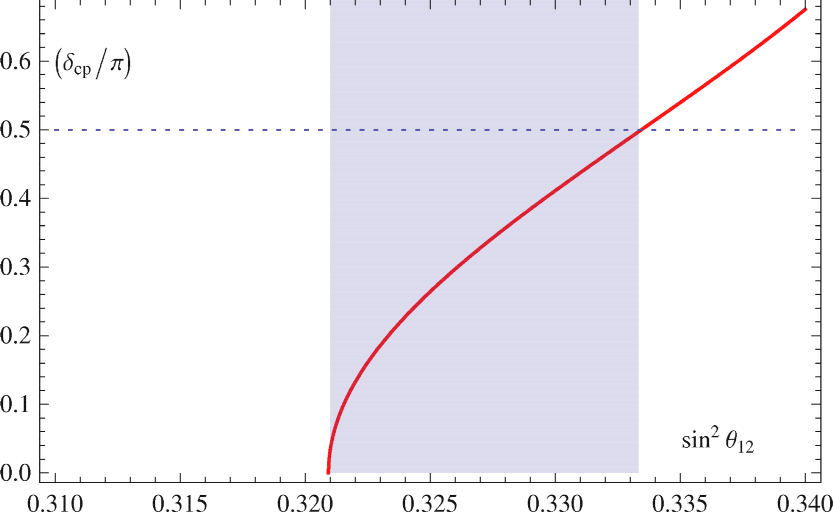} 
\caption{\footnotesize On introducing the Dirac CP phase $\delta_{cp}$ in the neutrino mixing matrix $U_{\nu}$ through $U_{\nu}^{c}$ and conducting charged lepton correction with $U_{eL}$, the prediction of $\sin^{2}\theta_{12}$ is affected. In order to achieve the best-fit of solar angle, $\delta_{cp}$ must go to zero. TBM mixing is recovered at $\delta_{cp}= \pi/2$. Also it can be noted nonzero values of $\delta_{cp}$ is possible in the neighbourhood (right side only) of the best fit ($0.32$) of $\sin^2\theta_{12}$.}
\end{center}
\end{figure}

We choose,
\begin{eqnarray}
\tilde{U}(\lambda, \delta_{cp})=\begin{pmatrix}
1 &\frac{1}{24}\lambda e^{-i\delta_{cp}} &-\frac{1}{24}\lambda e^{-i\delta_{cp}} \\ 
-\frac{1}{24}\lambda e^{i\delta_{cp}} & 1 & 0\\ 
 \frac{1}{24}\lambda e^{i\delta_{cp}} & 0  & 1 
\end{pmatrix}.
\end{eqnarray}
The $U_{\nu}(\lambda)$ becomes,
 \begin{eqnarray}
U_{\nu}(\lambda,\delta_{cp})\simeq \begin{pmatrix}
\sqrt{\frac{2}{3}}+\frac{\lambda e^{-i \delta_{cp}}}{12 \sqrt{6}} &\frac{1}{\sqrt{3}}-\frac{\lambda e^{-i \delta_{cp}}}{12 \sqrt{3}}  & 0 \\ 
-\frac{1}{\sqrt{6}}+\frac{\lambda e^{i \delta_{cp}}}{12 \sqrt{6}} &\frac{1}{\sqrt{3}}+\frac{\lambda e^{i \delta_{cp}}}{24 \sqrt{3}}  & \frac{1}{\sqrt{2}} \\ 
\frac{1}{\sqrt{6}}-\frac{\lambda e^{i \delta_{cp}}}{12 \sqrt{6}} & -\frac{1}{\sqrt{3}}-\frac{\lambda e^{i \delta_{cp}}}{24 \sqrt{3}} & \frac{1}{\sqrt{2}}
\end{pmatrix} 
\end{eqnarray} 
Similarly, on investigating the PMNS matrix  $U_{PMNS}$ on same footing as earlier, the effect of $\delta_{cp}$ is realized only in the solar angle prediction; other angles remain same as before[eqs.(\ref{s1},\ref{s3})],
\begin{eqnarray}
\label{s22}
\sin^{2}\theta_{12}=\frac{1}{3}-\frac{\lambda}{18}(\cos\delta_{cp}-\frac{\lambda}{24}).
\end{eqnarray}
From the above analysis this is clear that $\delta_{cp}$ remains arbirary if it resides in lepton sector. But the scenario is different if Dirac CP phase enters through neutrino sector in the sense that prediction of solar angle is affected. Comparing eqs.(\ref{s2},\ref{s11},\ref{s22}) it can be inferred that $\delta_{cp}$ must go to zero if we want $\sin^{2}\theta_{12}=0.321$. But interestingly, nonzero $\delta_{cp}$ with its value in the range $0<\delta_{cp}<\pi/2$ are possible for $\sin^2\theta_{12}$ in the range $0.32<\sin^2\theta_{12}<0.33$ [Fig.\ref{dd}].

\section{Discussion}
Corrections to TBM mixing has been implemented from both charged leptonic sector as well as from neutrino sector. This is different from the works where corrections are conducted either from anyone of the two\cite{cg,hm,ad,ll,ys,nq1,em7,sm,sd,ps}. Our approach finds some similarity with the ref.\cite{yh}. But the correction matrix used by the author is differtent from what we derived in the eq.(\ref{uc}).

We derived the diagonalizing matrix $U_{\nu}$, in eq.(\ref{unu}) starting from parametrization of a $\mu-\tau$ symmetric neutrino mass matrix following NH mass pattern, $M_{\mu\tau}$ along-with two hypothesis that mass-squared differeneces can be expressed in terms of Wolfenstein parameter $\lambda\sim 0.2253$. In section\ref{pmmt}, we have generated the neutrino mixing matrix $U_{\nu}(\lambda)$ starting from a $\mu-\tau$ symmetric neutrino mass matrix,  giving rise to mass eigenvalues with normal spectrum. One question arises  why out of the two mass hierarchies, we have preferred one than the others? In fact there is no strong reason behind that. But in ref\cite{csl2}, the author showed that the $2-3$ symmetry present in the texture of the neutrino mass matrix in eq.(\ref{mmutau}) is possible only with NH hierarchy of the neutrino mass spectrum. Also in ref.\cite{kh}, the analysis done by the author, based on ongoing CMB observations, including B mode polarization hints for NH pattern. The model within the $\mu-\tau$ symmetry regime, is expressible with two important parameters $\lambda$ and $\eta$ (flavour-twister). The model shows a strong preference towards the choice $\eta=\lambda$ which leads to the close agreement of solar mass squared difference, $\Delta m^2_{21}\,(\eta,\lambda)$ and solar mixing angle, $\theta_{12} (\eta)$ to the best-fits.

The charged lepton diagonalizing matrix, $U_{eL}$ we constructed in
eq.(\ref{uel}), does not affect the solar angle further, lowers the
atmospheric and produces nonzero reactor angles, when combined with
$U_{\nu}$ in order to give $U_{PMNS}$. The results are summarised as :
$\sin^2 \theta_{12}= 0.321$, $\sin^2\theta_{23}=0.425$ and
$\sin^{2}\theta_{13}=0.025$. Our analyses shows that if Dirac CP phase
resides in the leptonic sector, does not affect the predictions of
mixing angles. On the contrary, if it emerges out of the neutrino
sector, it affects only the prediction of $\theta_{12}$. Nonzero
values of $\delta_{cp}$ within the range $0-\pi/2$ are possible,
 and this corresponds to  $\sin^2\theta_{12}$ ranging from $0.321-0.333$. This is to be emphasised that $\sin^2\theta_{12}=0.321$ is possible only when $\delta_{cp}=0$. 

With $\theta_{12}^{CKM} \approx 13.02^0$ and $\theta_{23}^{CKM} \approx 2.35^0$\cite{nk}, we try to investigate the quark-lepton complementarity (QLC) and selfcomplementarity (SLC)\cite{ayu,mr,yj,xz}. We take $\theta^{PMNS}_{12}= 34.51^{0}$, $\theta^{PMNS}_{13}=9.2^{0}$ and  $\theta^{PMNS}_{23}=40.69^{0}$ as per the prediction of the model and find 
\begin{eqnarray}
\theta_{12}^{CKM}+\theta_{12}^{PMNS}&=& 45^0 +\delta \theta,\quad\quad \quad\text{(QLC)}\nonumber \\
\theta_{23}^{CKM}+\theta_{23}^{PMNS}&=& 45^0 -\delta \theta,\quad \quad\quad \text{(QLC)}\nonumber\\
\theta_{12}^{PMNS}+\theta_{13}^{PMNS}&=& \theta_{23}^{PMNS} -\delta \theta.\quad \: \: \: \text{(SLC)}\nonumber.
\end{eqnarray} 
where $\delta \theta\approx 2^0$, represnts the amount of deviation from the original QLC and SLC relations. This is interesting to note that all the three relations experience similar deviation $\delta \theta$ and also $\delta \theta \sim \theta^{CKM}_{23}$. One way to infer from the above results is either $\delta\theta$ is the error or the original three relations described above (with $\delta\theta=0$), need little modification. This is due to the fact that $\delta\theta$ and $\theta_{23}^{CKM}$ are equivalent. We hope that these simultaneous similar deviations are not unnatural and by replacing $\delta\theta$ with $\theta_{23}^{CKM}$ we try to reformulate the original three phenomenological relations. Perhaps,
\begin{eqnarray}
\theta_{12}^{CKM}-\theta^{CKM}_{23}+\theta_{12}^{PMNS}&=& 45^0,\nonumber \\
2\theta_{23}^{CKM}+\theta_{23}^{PMNS}&=& 45^0,\nonumber\\
\theta_{23}^{CKM}+\theta_{12}^{PMNS}+\theta_{13}^{PMNS}&=& \theta_{23}^{PMNS},\nonumber
\end{eqnarray} 
could be better QLC representations than the earlier. We hope that this little modification is permissible in the sense that $\theta_{23}^{CKM}$ is very small in comparison to the other members present in the above relations.


\begin{thebibliography}{1}
\bibitem{pfh}P.F Harrison, D.H Perkins and W.G Scott, Phys.Lett.\textbf{B530} 167 (2002), hep-ph/0202074.
\bibitem{em1}E. Ma and G. Rajasekaran, Phys.Rev.D \textbf{64},(2001)113012, hep-ph/0106291.
\bibitem{em2}E. Ma, Mod.Phys.Lett.A \textbf{17} (2002), 627-630, hep-ph/0203238.  
\bibitem{em3}E. Ma, Phys.Rev.D\textbf{70},031901(2004),hep-ph/0404199
\bibitem{em4}E. Ma, Phys.Rev.D \textbf{73},057304(2006), hep-ph/0511133.
\bibitem{em5}E. Ma, Phys.Lett.B \textbf{632}, 352 (2006), hep-ph/0508231.
\bibitem{em6}E. Ma, Mod.Phys.Lett.A \textbf{20}, 2601(2005), hep-ph/0508099.
\bibitem{a1}G. Altarelli and F. Feruglio, Nucl.Phys.B \textbf{720},64(2005),hep-ph/0504165.
\bibitem{a2}G. Altarelli and F. Feruglio, Nucl.Phys.B \textbf{741},215(2006), hep-ph/0512103.
\bibitem{az}A. Zee, Phys.Lett.B \textbf{630},58(2005), hep-ph/0508278.
\bibitem{ba}B. Adhikary and A. Ghosal, Phys.Rev.D \textbf{75},073020(2007),hep-ph/0609193.
\bibitem{csl1}C. S. Lam, Phys.Rev.D \textbf{78},073015(2008), hep-ph/0809.1185.
\bibitem{wg}W. Grimus, L. Lavoura and P. O. Ludl, J.Phys.G \textbf{36},115007 (2009),hep-ph/0906.2689.
\bibitem{fbl1}F. Bazzocchi, L. Merlo and S. Morisi, Nucl.Phys.B \textbf{816},204 (2009), hep-ph/0901.2086.
\bibitem{fbl2}F. Bazzocchi, L. Merlo and S. Morisi, Phys. Rev. D \textbf{80},053003 (2009), hep-ph/ 0902.2849.
\bibitem{tk1}T. Kobayashi, S. Raby and R. J. Zhang, Nucl.Phys.B \textbf{704},3(2005),hep-ph/0409098.
\bibitem{tk2}T. Kobayashi, H. P. Nilles, F. Ploger, S. Raby and M. Ratz, Nucl.Phys.B \textbf{768},135(2007),hep-ph/0611020.
\bibitem{ht}H. Ishimori, T. Kobayashi, H. Ohki, H. Okada, Y. Shimizu and M. Tanimoto, Prog.Theor.Phys.Suppl.183 (2010)1-163, hep-ph/1003.3552.
\bibitem{wk}L. Wolfenstein, Phys.Rev.Lett. \textbf{51} (1983) 1945.
\bibitem{nk}K. Nakamura et al.(Particle Data Group), J.Phys.G \textbf{37},(2010) 075021.
\bibitem{dc}DOUBLE-CHOOZ Collaboration, Y.Abe et al.,Phys.Rev.Lett.\textbf{108},131801 (2012),\textbf{arXiv:1112.6353}[hep-ex].
\bibitem{db}DAYA-BAY Collaboration,F.An et al.,Phys.Rev.Lett.\textbf{108},171803 (2012), \textbf{arXiv:1203.1669}[hep-ex].
\bibitem{rn}RENO Collaboration, J. Ahn et al.,Phys.Rev.Lett.\textbf{108},191802 (2012),\textbf{arXiv: 1204.0626}[hep-ex]. 
\bibitem{tk}T2K Collaboration, K. Abe et al., Phys.Rev.Lett.\textbf{107}, 041801 (2011), \textbf{arXiv: 1106.2822}[hep-ex].
\bibitem{pt}R. Nichol, Plenary talk at the Neutrino 2012 conference, http: //neu2012.kek.jp/.
\bibitem{gf}G. Fogli et al., Phys.Rev.\textbf{D86},(2012) 013012, \textbf{arXiv:1205.5254}.
\bibitem{mg}M. Gonzalez-Garcia, M. Maltoni and J. Salvado, JHEP \textbf{1004}, 056 (2010), \textbf{arXiv: 1001.4524}.
\bibitem{bp}B.Pontecorvo, Sov.Phys. JETP 26, 984(1968); Z. Maki, M. Nakagawa, and S. Sakata, Prog.Theor.Phys.28,870 (1962).
\bibitem{sk} S.F King, Phys.Lett.B \textbf{718} (2012),136, hep-ph/ 1205.0206.
\bibitem {vl}D.V. Forero, M. Tortola and J.W.F Valle, Phys.Rev.D \textbf{86}(2012) 073012, \textbf{arXiv: 1205.4018v4}.
\bibitem{csl2}C.S Lam, Phys.Lett.B\textbf{507}(2001),214-218, hep-ph/0104116.
\bibitem{nh}N. Haba, A. Watanbabe and K Yoshioka, Phys.Rev.Lett.\textbf{97},(2006)041601, hep-ph/0603116.
\bibitem{nns1}N. Nimai Singh, M. Rajkhowa and A Borah, J.Phys.G34(2007)345-352, hep-ph/0603154.
\bibitem{nns2}C. Duarah, A. das and N. Nimai Singh, Phys.Lett.B\textbf{718} (2012) 147-152, hep-ph/ 1207.5225.
\bibitem{cg}C. Giunti and M. Tanimoto, Phys.Rev.D \textbf{66}(2002)053013, hep-ph/0207096.
\bibitem{hm}H. Minakata and A. Y. Smirnov, Phys.Rev.D \textbf{70}(2004)073009, hep-ph/0405088.
\bibitem{ad}A. Datta, L. Everett and P. Ramond, Phys.Lett.B\textbf{620}(2005)42, hep-ph/0503222.
\bibitem{ll}L. L. Everett, Phys.Rev.D\textbf{73}(2006)013011, hep-ph/0510256.
\bibitem{ys}Y. Shimizu, M. Tanimoto and A. Watanabe, Prog.Theor.Phys.\textbf{126},81(2011), hep-ph/ 1105.2929.
\bibitem{nq1}N. Qin and B. Q. Ma, Phys.Lett.B\textbf{702}, 143 (2011), hep-ph/1106.3284.
\bibitem{em7}E. Ma and D. Wegman, Phys.Rev.Lett.\textbf{107},061803 (2011), hep-ph/ 1106.4269.
\bibitem{sm}S. Morisi, K. M. Patel and E. Peinado, Phys.Rev.D\textbf{84},053002 (2011), hep-ph/1107.0696.
\bibitem{sd}S. Dev, S. Gupta and R. R. Gautam, Phys.Lett.B \textbf{704},527 (2011), hep-ph/1107.1125.
\bibitem{ps}P. S. Bhupal Dev, R. N. Mohapatra and M. Severson, Phys.Rev.D \textbf{84}, 053005 (2011), hep-ph/ 1107.2378.
\bibitem{yh}Y. H Ahn, H. Cheng and S. Oh, Phys.Rev.D\textbf{84} (2011) 113007, hep-ph/1107.4549.
\bibitem{kh}K Hirano, arXiv:1212.6423[astro-ph.CO].
\bibitem{ayu}A. Yu. Smirnov, hep-ph/ 0402264.
\bibitem{mr}M. Raidal, Phys.Rev.Lett.\textbf{93}, 161801 (2004), hep-ph/ 0404046.
\bibitem{yj}Y.j. Zheng and B.Q. Ma, Eur.Phys.J.Plus \textbf{127},7(2012), hep-ph/1106.4040.
\bibitem{xz}X. Zhang and B.Q. Ma, Phys.Lett.B \textbf{710}, 630 (2012) hep-ph/1202.4258.
\end{thebibliography}
\end{document}